\documentclass[aps,prb,reprint,showpacs,floatfix,amsmath,amssymb,
showkeys,showpacks]{revtex4-1}

\usepackage{graphicx}
\usepackage{epstopdf}
\usepackage{color}
\usepackage{hyperref}
\usepackage{bm}

\newcommand{\CuSe}{{\text{Cu}_2 \text{O} \text{Se} \text{O}_3}}

\renewcommand{\vec}[1]{\mathbf{#1}}

\newcommand{\ka}{{\mathbf k}}

\newcommand{\qu}{{\mathbf q}}
\newcommand{\Qu}{{\mathbf Q}}

\newcommand{\ra}{{\mathbf r}}

\newcommand{\no}{\noindent}

\newcommand{\bra}[1]{ \langle #1 |}
\newcommand{\ket}[1]{|#1  \rangle}
\newcommand{\expect}[1]{ \langle #1  \rangle}

\begin{document}
\title{Skyrmion Lattices in Electric Fields}
\author{Alex \,J.\, Kruchkov}
\email{alex.kruchkov@epfl.ch}
\author{Henrik M.\, R{\o}nnow}
\affiliation{Laboratory for Quantum Magnetism (LQM), {\'E}cole Polytechnique F\'{e}d\'{e}rale de Lausanne (EPFL),  Station 3, CH-1015 Lausanne, Switzerland}

\date{\today}

\begin{abstract}
This paper studies the influence of electric fields on the skyrmion lattice (SkL) in insulating skyrmion compounds with weak magnetoelectric (ME) coupling. The ME coupling mechanism is an interaction between the external electric field $E$ and local magnetization in the sample. Physically, the $E$-field perturbs the spin modulation wave vectors resulting in the distortion of the SkL and the $E$-field induced shift in energy. Due to the relativistic smallness of ME coupling, the important physics is captured already in the elastic ($\propto$$E$) and inelastic ($\propto$$E^2$) responses. In this spirit, the effect of the fourth-order cubic anisotropy responsible for stabilization of the skyrmion phase is taken into account perturbatively.  
The shift in energy can be either positive or negative, -- depending on the direction of electric field, -- thus stabilizing or destabilizing the SkL phase. 
Understanding the $E$-field energetics is important from the viewpoint of creation (writing) and destruction (erasing) of skyrmion arrays over the bulk, which is paramount for developing skyrmion-based logical elements and data storage.
\end{abstract}


\maketitle

%
%
%
%



\section*{Introduction}

Magnetic skyrmions are quasiparticles built from topologically protected vortices of spins in helimagnets.  They were theoretically predicted by Bogdanov in 1989  as  thermodynamically stable states of chiral magnets,\cite{Bogdanov1989} and discovered twenty years after in the form of a  hexagonal skyrmion lattice in the helimagnetic conductor MnSi.\cite{Binz2006,Muhlbauer2009} The magnetic skyrmions gained their name after T. Skyrme, who introduced such field configurations in the context of low-energy physics of mesons and baryons \cite{Skyrme1961} to explain the stability of particles through topological protection against continuous field transformations. 
Currently, skyrmions in chiral magnets are compelling due to their nanoscale size and the mentioned topological protection: They are envisaged as promising information carriers, and the skyrmion racetrack memory was recently proposed, see e.g. Refs.  
\cite{Fert2013,Nagaosa2013,Racetrack1,Racetrack2,Racetrack3}

As quasiparticles, skyrmions can form a crystalline state, and  solid-state concepts, such as symmetry breaking, order parameter, elementary excitations etc., can be applied, therefore opening doors for simple but very efficient theoretical models. 
The skyrmion crystal was observed both in the reciprocal and real space  in the form of a two-dimensional hexagonal lattice,
\cite{Muhlbauer2009,Tonomura2012} as is sketched in Figure 1.
The hallmark of the skyrmion lattice (SkL) as seen by SANS (small angle neutron scattering) is the appearance of a six-fold pattern in reciprocal space.\cite{Muhlbauer2009,Munzer2010,Tokunaga2015,Adams2012,White2014}  
The thermodynamical stability of skyrmion lattices is  improved experimentally by continuous tuning of coupling parameters with hydrostatic pressure \cite{Levatic2016} or by perturbing the skyrmion lattice with electric fields. \cite{Okamura2016}
The interest in the skyrmion lattices and their stability is justified by the solid-state concept that the lattice could be melted into individual skyrmions.


\begin{figure}[b]
	\includegraphics[width=0.9 \columnwidth]{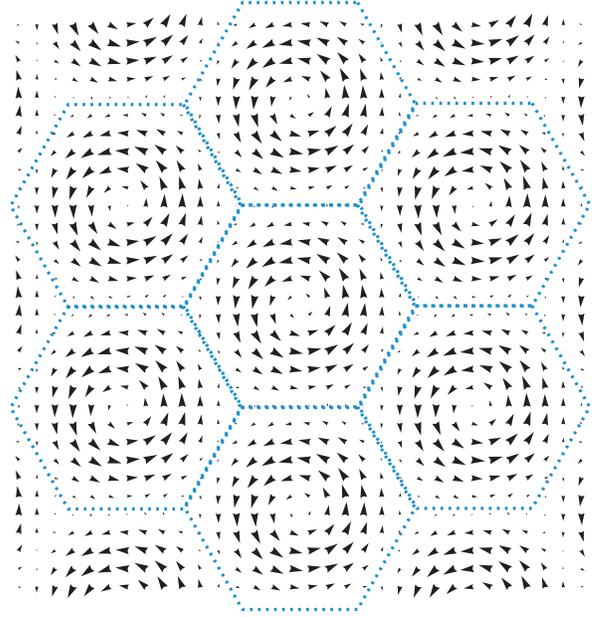}
	\caption{The hexagonal skyrmion lattice (real space), as a schematic projection of spins on the basal plane. The dotted lines (blue) denote the in-plane periodicities of the SkL vector order parameter. }
\end{figure}


Manipulation and control of skyrmions have become an active topic of skyrmionics. 
Recent experiments succeeded  manipulation of skyrmions with  moderate electric fields, electric currents
 , and thermal gradients. \cite{White2012,White2014,Jonietz2010,Yu2012,Jiang2015,Everschor2012,Mochizuki2014,Watanabe2016}
To avoid the Ohmic heating effects which are undesirable for electronics, 
the application of moderate electric fields to insulating skyrmion-host compounds (such as $\CuSe$ \cite{Seki2012}) is potentially more advantageous for the current-driven devices.  
These observations motivated theoretical proposals for creation (``writing'') skyrmions in insulating helimagnets with the help of external electric field, \cite{Mochizuki2015,Mochizuki2016,Okamura2016,Watanabe2016}  and subsequent electric-field guiding. \cite{Upadhyaya2015}
To date, a key experimental challenge in this direction is stabilization and control of the skyrmion lattice by an electric field. 
Therefore, there is an urge for simple theories of skyrmion lattice response to the E-field. 
In particular, what is of interest is the shift of skyrmion lattice energy in electric field, with subsequent stabilization of the skyrmion lattice.\cite{Okamura2016}

In the present paper we discuss the physics of the skyrmion lattice upon application of $E$-field to an insulating skyrmion-host compound, such as $\CuSe$. 
In this material, the effect of  magnetoelectric coupling arises due to a hybridization mechanism originating from relativistic spin-orbit interaction (see Refs. \cite{Seki2012,Liu2013,Nagaosa2007,Belesi2012}), which gives rise to an electric dipole moment which can be expressed in terms of the local spin variables. The strength of the effect is however relativistically small, which allows us to build an accurate perturbation theory in $E$-fields.

 In this paper, we treat the effect of electric field on skyrmion lattices in the two first orders of perturbation theory. The skyrmion lattice in electric fields becomes slightly distorted, and we introduce the elastic and inelastic distortion vectors. The shift of SkL energy comes from expectation values of the electromagnetic coupling and anisotropic contributions.

This paper is organized as follows. 
In Sec. I, we write down the effective coarse-grained energy functional, and make a rotation of quantization axes, as demanded by the experimental orientation of magnetic field. 
In Sec. II, we consider the mean-field treatment for the SkL energy density with the magnetic field applied. 
In Sec. III, we calculate the distortion of the skyrmion lattice by the electric field in an elastic approximation.
In Sec. IV, we first calculate the mean-field expectation value of the electromagnetic coupling term, also taking into account the distortion-induced anisotropic contributions. To be consistent in the first two orders of perturbation expansion, we introduce the inelastic distortion vector, and write down all the terms in first and second orders in dimensionless electric field $\ae$. In the concluding section  we highlight the main results and discuss the limitations of the model and possible applications. 


%
%
%
%

\section{Energy density in a coarse-grained model}

The Skyrmion Lattice (SkL) is a long-range-order spin configuration which can be visualized as a hexagonal array of vortices (see Fig. 1). Experimentally, the hallmark of the SkL phase is appearance of a six-fold reflection pattern in reciprocal space, as sketched in Fig. 2, each of the wave vectors are rotated by $2 \pi/3$ (see e.g. Refs. \cite{Muhlbauer2009,White2014} for SANS patterns). 
In this study, we describe the skyrmion lattice by a coarse-grained magnetization $\vec S (\ra)$, which can be built on the three $\vec Q$-vectors (Fig. 2). With a good accuracy,\cite{Muhlbauer2009,Binz2008}
 the SkL phase can be approximated by the multispiral spin structure,

\begin{equation}\begin{split}\begin{gathered}
\label{SkL order parameter}
\vec S (\ra)  = \vec m   
+ \mu 
\sum_{\vec Q_n}  \vec S_{\vec Q_n} e ^{i \vec Q_n \ra+i \varphi_n} 
+  \vec S^{*}_{\vec Q_n} e ^{-i \vec Q_n\ra-i \varphi_n} ,
\end{gathered}\end{split}\end{equation}

\no
where $\vec m \equiv \langle \vec S (\ra ) \rangle$ is a uniform magnitization,  with (spatial) average defined as $\langle... \rangle = \int \frac{dV}{V} (...)$ throughout the study, and $\mu$ is the weight of the SkL helical modulations. The sum in \eqref{SkL order parameter} runs over the ``3Q-structure'' (Fig. 2), 
the  relative phases $\varphi_n$  in \eqref{SkL order parameter} are important for minimization of the SkL energy.
The expectation of energy density in the coarse-grained model is given by calculating the spatial average $\langle \mathcal{H} \rangle$ with spin function


\begin{figure}[t]
	\includegraphics[width=0.95 \columnwidth]{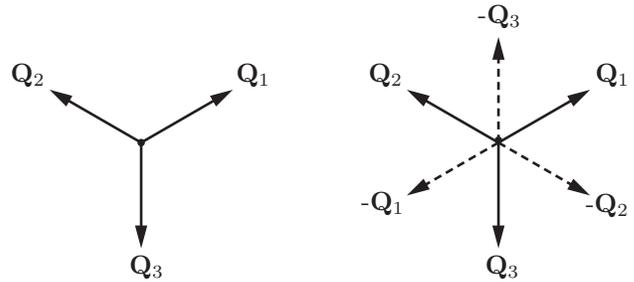}
	\caption{The multispiral ("3Q") SkL structure in reciprocal space. Right-hand side of the figure shows the auxiliary negative reflexes, so that the hexagonal real-space  SkL is built on the six wave vectors  $\{ \vec Q_1, \text{-} \vec Q_3 , \vec Q_2, \text{-} \vec Q_1, \vec Q_3, \text{-} \vec Q_2 \}$, see the main text.} 
\end{figure}


\begin{equation}
\mathcal{H} =
 \mathcal{H}_{JDh}
 +
  \mathcal{H}_{A}
  +
 \mathcal{H}_{\alpha E},
 \end{equation}

\no
where the helimagnetic term

\begin{equation}
\label{H JDh}
 \mathcal{H}_{JDh}=J(\nabla \vec{S})^{2}+D\vec{S}\cdot(\nabla \times \mathbf{S})-\mathbf{h}\cdot \vec S 
 \end{equation}
 
\no
takes into account Heisenberg interaction ($J$), Dzialoshinskiy-Moriya interaction ($D$) and the Zeeman coupling to the external magnetic field  $\vec h$.
This coarse-grained model works 
when the following hierarchy of energies is respected: (1) the strongest is Heisenberg exchange parameter $J$ which favors the ferromagnetic alignment; 
(2) the Dzialoshinskiy-Moriya interaction (DMI) is slightly tilting two adjacent spins thus resulting helical modulations. 
(3) The cubic magnetocrystalline anisotropy $\mathcal H_A$ is considered the weakest in this hierarchy.
Finally, 
the weak magnetoelectric term $\mathcal{H}_{\alpha E}$ is perturbatively small  as estimated further.
The spatial modulation of the ordered phase (the wavelength of the helices in the ground state) is given by a wave length of order $\lambda /a  \sim J/D \gg  1$, i.e. much larger than the crystal lattice parameter $a$: for example, in $\CuSe$, $\lambda =630 \, \text{\AA}$, $a =8.9 \, \text{\AA} $, which gives $\lambda /a \sim 70$. 
In such case, the magnetization on neighboring lattice sites is varying very slowly and the physics of the system in the ordered phase is appropriately described by a continuous-limit model as assumed in \eqref{H JDh}.

In this study, we consider the fourth-order anisotropy as it represents the essential physics of the problem by stabilizing the SkL phase.\cite{White2014,Muhlbauer2009}
The symmetry of $\CuSe$ is described by the $P2_1 3$ space group, which allows a fourth-order magneto-crystalline anisotropy of the form 
$A_1 (S_{x}^{4}+S_{y}^{4}+S_{z}^{4}) 
+ A_2 (S_x^2 S_y^2+S_y^2 S_z^2+S_z^2 S_x^2 )$, 
which can be reduced to

\begin{equation}
\label{anisotropy1}
\mathcal{H}_{A}=A(S_{x}^{4}+S_{y}^{4}+S_{z}^{4}).
\end{equation}

\no
Indeed, proceeding to the unitary  parametrization $\vec S/|\vec S|=(\sin \theta \cos \psi, \sin \theta \sin \psi, \cos \theta)$, 
one obtains $S_x^2 S_y^2+S_y^2 S_z^2+S_z^2 S_x^2 =
- \frac{1}{2} (S_{x}^{4}+S_{y}^{4}+S_{z}^{4})  + \frac{1}{2}$, thus $A = A_1 - A_2/2$.

The magneto-electric coupling arises due to the weak  $p$-$d$ hybridization mechanism(see Refs. \cite{Seki2012,Liu2013,Nagaosa2007,Belesi2012}), which gives rise to an electric dipole moment $\vec P = \alpha_{\lambda} (S_y S_z, S_z S_x, S_x S_y)$, i.e. the {\textit{electric dipole moment}} coupled to the spin variables $S_x,S_y,S_z$.  Therefore, in external electric fields the ordered phase is perturbed by  $-\vec P \cdot \vec E$, or

\begin{equation}
\label{H ae}
\mathcal{H}_{\ae}=
\alpha E_{x}S_{y}S_{z}+ \text{cyclic permutations} ,
\end{equation}

\no
where $\vec E = (E_x, E_y,E_z)$ is the external electric field and for simplicity we absorbed the minus sign into $\alpha = - \alpha_{\lambda}$. 
For $\CuSe$ the strength of magneto-electric coupling is estimated as $|\alpha|  \sim 10^{-33} \, \text{Jm}/\text{V}$, see Ref.\cite{Omrani2014}.

The above expressions \eqref{H JDh}-\eqref{H ae} are written in the natural frame. Experimentally, one often needs to apply magnetic and electric fields along directions where particular properties of the system are better revealed. In particular, in some $E$-field rotation experiments,\cite{White2014} the magnetic field $\vec h$ is parallel to
$[1 \, \bar 1 \, 0]$, while the electric field is parallel to $[1 1 1]$ or $[\bar 1 \bar 1 \bar 1]$, with 
$|E_x|=|E_y|=|E_z| \equiv E$ for simplicity of notations (thus the magnitude is $|\vec E| = \sqrt 3 E$). It is easier to carry out calculations in the rotated spin frame $(x',y',z')$, with $\hat z'$ set by the direction of magnetic field $\vec h$, see Fig. 3.  
In the case of above-mentioned geometry (Fig. 3), the transformation of the rotated spin frame is given by the rotation matrix

\begin{equation}\begin{split}\begin{gathered}
\label{rotated frame}
\mathcal{R}
=
\frac{1}{\sqrt{2}}
 \left( \begin{array}{ccc}
1  & 0 & 1   \\
1  & 0 & {\text-} 1   \\
0 & \sqrt{2} & 0 
\end{array} \right).
\end{gathered}\end{split}\end{equation}


\begin{figure}
\begin{center}
	\includegraphics[width= 0.45 \columnwidth ]{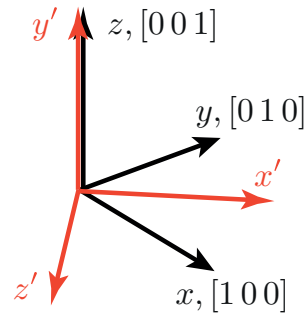}
	\end{center}
	\caption{Rotated spin frame (red) for the magnetic field orientation $[1 \, \bar 1 \, 0]$.}
\end{figure}


\no
While the helimagnetic term \eqref{H JDh} is a Lifshitz invariant, the fourth-order anisotropy \eqref{anisotropy1} is transformed under rotation \eqref{rotated frame} into

\begin{equation}\begin{split}\begin{gathered}
\label{anisotropy2}
 \mathcal{H}_{A}=A\left(\frac{1}{2}S_{x'}^{4}+S_{y'}^{4}+3S_{x'}^{2}S_{z'}^{2}+\frac{1}{2}S_{z'}^{4}\right),
\end{gathered}\end{split}\end{equation}

\no
The $E$-field perturbation \eqref{H ae} also transforms under rotation ${\mathcal R}$,

\begin{equation}\begin{split}\begin{gathered}
\label{EP coupling}
 \mathcal{H}_{\ae} =\frac{\alpha E}{2}\left(S_{x'}^{2}+2 \sqrt{2}S_{x'}S_{y'}-S_{z'}^{2}\right), 
\end{gathered}\end{split}\end{equation}

\no
and the electric field is directed along $\hat e_{x'} + \hat e_{y'}$. 
For simplicity, we drop  the prime signs in the subsequent calculations.

We make a remark on the importance of the fourth-order magnetocrystalline anisotropy \eqref{anisotropy1}. The appearance of the $3 \Qu$-structure with    $\Qu_1 + \Qu_2 + \Qu_3 = 0$  with helices phased in a way to form the two-dimensional skyrmion crystalline occur only due to higher-order energy terms represented in the model.
In chiral ferromagnets, the anisotropy of at least the fourth order can be considered as the source of the SkL order parameter. Indeed, we can consider the fourth order anisotropy  which contains - fully or partially - the term
$\vec S^4(\ra) = S_{x}^{4}+S_{y}^{4}+S_{z}^{4} + 2(S_x^2 S_y^2+S_y^2 S_z^2+S_z^2 S_x^2)$. Introducing now  new variable without uniform magnetization $\vec s(\vec r) = \vec S (\vec r) - \vec m$, the fourth-order term  will contain a cubic term

\begin{equation}\begin{split}\begin{gathered}
\label{cubic term}
\vec S^4 (\ra)  = \left[  \vec s (\ra) + \vec m \right]^4 = 
...
+
4 \vec s^2  (\ra)  \, \vec s  (\ra) \cdot \vec m + ... ,
\end{gathered}\end{split}\end{equation}

\no
where we have not mentioned the other terms (linear, quadratic, quartic). The spatial average of the cubic term in the left-hand side of \eqref{cubic term}
can be Fourier-transformed as

\begin{align}
\label{cubic invariant}
\vec m \expect{ \vec s^2  (\ra)  \, \vec s  (\ra)  } 
& = 
 \sum_{\vec k_{1,2,3}} 
(\vec S_{\ka_1} \cdot \vec S_{\ka_2} ) ( \vec m \cdot \vec S_{\ka_3} ) e^{i (\varphi_{\ka_1}+\varphi_{\ka_2}+\varphi_{\ka_3})}
\\
&
\times \mu^3\,  \delta(\ka_1+\ka_2+\ka_3),
\end{align}

\no
where $S_{\ka}$ and their phases are defined as in Eq.\eqref{SkL order parameter}, and $\ka_{1,2,3}$ can be in principle any wave vector.  If we define parameters in such a way that $\mu>0$, $(\vec S_{\ka_1} \cdot \vec S_{\ka_2} ) (\vec m \cdot \vec S_{\ka_3}) > 0 $, the hexagonal phase will be minimized if only $\varphi_{\ka_1}+\varphi_{\ka_2}+\varphi_{\ka_3} = \pi$ and $\ka_1+\ka_2+\ka_3 = 0$. The latest condition gives the so called $3 \vec Q$ structure, $\Qu_1 + \Qu_2 + \Qu_3 = 0$, with the three wave vectors equirotated by $2 \pi/ 3$ due to symmetry.\cite{Muhlbauer2009,Binz2006} This situation is shown  on Fig. 1.


%
%
%
%

\section{Skyrmion lattice in zero electric field: mean-field treatment}

In this section we consider the mean-field treatment of the skyrmion lattice. First, we find the single-helix eigenstates, which give rise to a modulated spin structures with wave length $\lambda = 2 \pi/ k_0 = 4 \pi J/D$. After that, we construct the skyrmion lattice order parameter, and calculate the mean-field energy of the skyrmion lattice.

We start from considering the interplay between the Heisenberg term and the DMI coupling,

\begin{equation}
\label{W0}
W_{0}=
\expect{
J \left[\nabla \vec{S(\ra)}\right]^2
 +D \, \vec{S} (\ra) \cdot \left[\nabla \times \vec {S} (\ra)\right] 
 }
= 
\sum_{\vec k} \vec S_{\ka}^{\dag}
\hat {\mathcal {H}}_0
  \vec S_{\ka},
\end{equation}

\no
where in we used the Fourier transform of the spatial average to reciprocal space. Here $\hat {\mathcal {H}}_0$ is  written in spin representation  $\vec S_{\ka} = (S_\ka^x, S_\ka^y, S_\ka^z)^{T}$ as a matrix operator

\begin{equation}\begin{split}\begin{gathered}
\label{H0 1}
\hat {\mathcal {H}}_0 = \left( \begin{array}{ccc}
J k^{2} & \text{-}iDk_{z} & iDk_{y} \\
iDk_{z} & J k^{2} & \text{-} iDk_{x} \\
\text{-} iDk_{y} & iDk_{x} & J k^{2} \end{array} \right). 
\end{gathered}\end{split}\end{equation}

\begin{figure*}
	\includegraphics[width= 170 mm ]{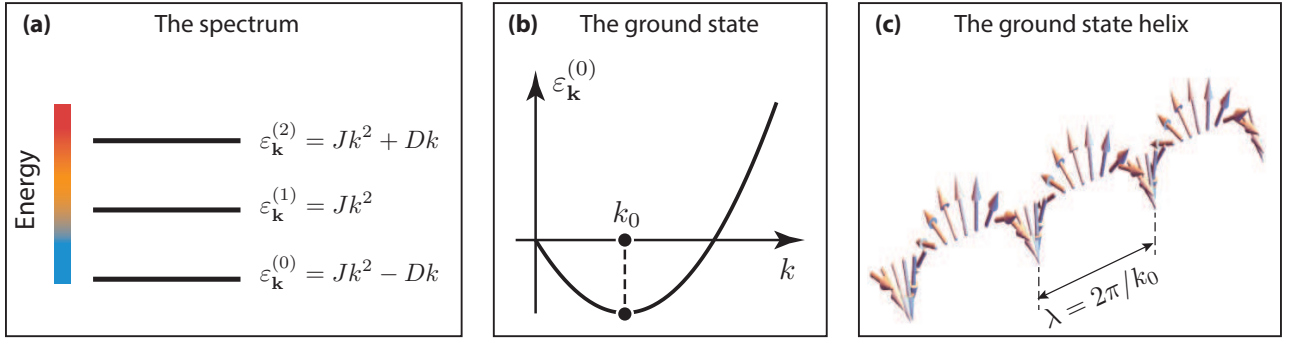}
	\caption{Appearance of the helical modulation: (a) The spectrum of $3\times3$ energy matrix $\hat {\cal{H}}_0$. (b) 
	Minimization of the state with lowest energy $\varepsilon^{(0)}_{\ka} =Jk^{2}-Dk$  gives helix wave vector  $k_0 = D/2J$.
	(c) Real-space visualization of the helix with $\lambda = 2 \pi/ k_0$. }
\end{figure*}

\no
The propagation vectors skyrmion lattice lie in the plane which is perpendicular to the magnetic field $\vec h$. Consequently, each of the six helices of the skyrmion lattice  is parametrized as $\ka = (k_z,k_y,0)$ in the rotated frame, thus the problem is effectively two-dimensional. The energy matrix ("hamiltonian") \eqref{H0 1}  is 

\begin{equation}\begin{split}\begin{gathered}
\label{H0 2}
\hat {\mathcal {H}}_0 = \left( \begin{array}{ccc}
J k^{2} & 0 & iDk_{y} \\
0 &J k^{2} & \text{-}iDk_{x} \\
\text{-}iDk_{y} & iDk_{x} & J k^{2} \end{array} \right) ,
\end{gathered}\end{split}\end{equation}

\no
and is diagonalized on the eigenstates

\begin{align}
\ket{ \vec S^{(0)}_{\ka}  } & 
=\frac{1}{\sqrt{2}}\left(\text{-}i\hat{k}_{y},i\hat{k}_{x},1\right)^{\intercal} ,
\label{Sk0} \\
\ket{ \vec S^{(1)}_{\ka}  } & 
=\left(\hat{k}_{x},\hat{k}_{y},0\right)^{\intercal} ,
\label{Sk1}
\\
\ket{ \vec S^{(2)}_{\ka}  } & 
=\frac{1}{\sqrt{2}}\left(i\hat{k}_{y},\text{-}i\hat{k}_{x},1\right)^{\intercal},
\label{Sk2}
\end{align}

\no 
where $\hat k_{x,y,z}= k_{x,y,z}/|\vec k|$ and we introduced "bra" and "ket" notations as shortcuts to write the perturbation formulas of the matrix mechanics \cite{Green1965} in a familiar way. Consistent with our previous notations, we denote $ \langle \vec S_{\vec k}| ... |\vec S_{\vec k} \rangle = \sum_{\vec k} \vec S_{\ka}^{\dag}
...
  \vec S_{\ka}$, which is just a Fourier-transform of  the corresponding spatial averaging $\langle... \rangle$ as in Eq.\eqref{W0}. 

The spectrum of matrix $\hat {\mathcal {H}}_0$ consists of the three equidistant energy solutions, with the energy separation $\pm D k$, see Fig. 4a,

\begin{align}
\varepsilon^{(0)}_{\ka}=Jk^{2}-Dk , 
\ \ \ \
\varepsilon^{(1)}_{\ka} =Jk^{2} , 
\ \ \  \
\varepsilon^{(2)}_{\ka}=Jk^{2}+Dk,
\end{align}

\no
For the positive $J$, $D$ the lowest energy solution is the helix

\begin{equation}\begin{split}\begin{gathered}
\label{ground state}
\ket{ \vec S^{(0)}_{\ka}  }
= \frac{1}{\sqrt{2}}
\left( \begin{array}{c}
\text{-}i \hat k_y \\
i \hat k_x \\
1 \end{array} 
\right),
\end{gathered}\end{split}\end{equation}

\no
which corresponds to the eigenvalue  $\varepsilon^{(0)}_{\ka}$. The modulation vector $k_0$ is determined by further minimization of the energy, 
${\partial \varepsilon^{(0)}_{\ka} } / {\partial k} = 0$. Consequently, the minimum of the dispersion $\varepsilon^{(0)}_{\ka}=Jk^{2}-Dk$  is satisfied if

\begin{align}
k_0 = \frac{D}{2 J},
\end{align}

\no
see also Fig. 3a,b. Finally,  it is instructive to rewrite one more time the eigenspectrum  of $\mathcal H_0$ as we use it in perturbation formulas further, 

\begin{align}
\label{spectr}
\varepsilon^{(0)}_{\ka_0}  = -  D k_0/2,
 \ \ \ 
\varepsilon^{(1)}_{\ka_0}  = D k_0/2, 
 \ \ \ 
\varepsilon^{(2)}_{\ka_0}  =  3 D k_0 /2.
\end{align}

\no
As a remark, it is usually convenient to measure the energy of the system in units of $D k_0$.

Next we consider the superposition of helices into a skyrmion lattice. 
Neglecting higher-order harmonics, the skyrmion lattice is approximated as uniform magnetization $m \hat e_z$ along magnetic field and a 3$Q$ multispiral configuration,

\begin{equation}
\label{SkL explicit}
\vec{S}(\ra) \simeq
\mu \sum_{i=1}^{3}
\left(
\vec S_{\vec Q_{i}}
e^{\vec Q_{i} \ra}
+ c.c. \right)
+ m  \, \hat{e}_z.
\end{equation}

\no
In the "braket" notation, the real-space skyrmion phase has a shortcut notation

\begin{equation}\begin{split}\begin{gathered}
\label{skl braket}
\ket{ \vec S (\ra)   } = 
\mu \sum_{\ka}^{\{ \pm \Qu_1...\pm \Qu_3\}}
\ket{ \vec S^{(0)}_{\ka}  }  e^{i \ka \ra+ i \varphi_{\ka} } 
+
 m \ket{ \vec S_{0} } ,
\end{gathered}\end{split}\end{equation}

\no
where $| \vec S^{(0)}_{\ka}  \rangle$ is given by Eq.\eqref{ground state} and the ferromagnetic component is $\ket{ \vec S_{0} } \equiv (0\, 0 \, 1)^{\intercal} $.


The skyrmion lattice possess some interesting properties. 
First, using the explicit form \eqref{SkL explicit}, the first-order moments are

\begin{equation}\begin{split}\begin{gathered}
\expect{  S_x (\ra) } = 0,
\ \ \ 
\expect{  S_y (\ra) } = 0,
\ \ \ 
\expect{  S_z (\ra) } = m,
\end{gathered}\end{split}\end{equation}

\no
so the uniform magnetization (the "ferromagnetic component")
$\vec m \equiv \expect{ \vec S(\ra) } = m \, \hat e_z $ 
is the only non-vanishing first-order moment. 
The second-order in-plane moments are

\begin{align}
\expect{  S_x^2 (\ra) }
& =
\sum^{\left\{ \qu_1...\qu_6 \right\} }_{\ka} 
 S^x_{\ka} S^x_{-\ka} = 3 \mu^2/2 ,
\\
\expect{  S_y^2 (\ra) }
& =
\sum^{\left\{ \qu_1...\qu_6 \right\} }_{\ka} 
 S^y_{\ka} S^y_{-\ka} = 3 \mu^2/2 ,
\end{align}

\no
while the longitudinal (z) second-order moment is  

\begin{equation}\begin{split}\begin{gathered}
\expect{  S_z^2 (\ra) }
=m^2+ 
\sum^{\left\{ \qu_1...\qu_6 \right\} }_{\ka} 
 S^z_{\ka} S^z_{-\ka} = m^2 + 3 \mu^2 .
\end{gathered}\end{split}\end{equation}

\no
One can verify that all the mixed moments vanish,

\begin{align}
\expect{  S_\alpha S_\beta }  = 0, 
\ \ \ 
\alpha \ne \beta,
\end{align}

\no
so, in general, one has

\begin{align}
\label{second moments}
\expect{  S_\alpha  S_\beta } = \frac{3}{2} \mu^2 \delta_{\alpha,\beta} [1+ (1+2m^2/3 \mu^2) \delta_{\alpha,z} ],
\end{align}

\no
where $\alpha, \beta={x,y,z}$. Formula \eqref{second moments} is handy for further expectation value calculations. 
For a fixed temperature one can use normalization $\expect{ \vec S^2 (\ra) }=1$, thus the following constraint holds\cite{White2014}

\begin{equation}
\label{constraint}
\expect{ \vec S^2 } = m^2 + 6 \mu^2 = 1.
\end{equation}

\no
Finally, it is interesting to note that in this approximation the fourth-order moment for the skyrmion lattice is surprisingly 
$\langle \vec S^4  \rangle  \ne \langle \vec S^2  \rangle^2$, i.e. the magnetization field is ``soft''. 

Now we consider the skyrmion lattice in finite magnetic fields by adding the source term  $-\vec h \cdot \vec S$,

\begin{equation}\begin{aligned}
\label{SkL energy}
W^{(h)}_{0}
&=
\expect{
  J \left[\nabla \vec{S(\ra)}\right]^2
 +D \, \vec{S} (\ra) \cdot \left[\nabla \times \vec {S} (\ra)\right] 
 - \vec{h} \cdot \vec S(\ra) 
}
 \\
& = 
\sum^{\left\{ \qu_1...\qu_6 \right\} }_{\vec k} 
\vec S_{\ka}^{\dag}
\
\hat {\mathcal {H}}_0
\
  \vec S_{\ka} - h \, m ,
\end{aligned} \end{equation}

\no
which however doesn't change the mean-field eigenstates \eqref{Sk0}-\eqref{Sk2} as it contains no spatial derivatives. 
Thus the expectation value \eqref{SkL energy} is modified only in terms of elongating the ferromagnetic component $m$, but the topology of the skyrmion order parameter is not affected. The mean-field treatment of the SkL energy \eqref{SkL energy} yields

\begin{equation}\begin{split}\begin{gathered}
\label{W0 in magnetic field}
W^{(h)}_{0}
= 
- 3 D k_0 \, \mu^2 - h \, m .
\end{gathered}\end{split}\end{equation}

\no
In the mean-field treatment,  one may minimize the total energy with respect to all the components of the magnetic moment, including $m$ and  $\mu$, if a comparison between two phases is needed. 
For example, 
minimizing now Eq.\eqref{W0 in magnetic field} with  constraint \eqref{constraint}, one gets the mean-field estimates  $m_{\text{MF}}=h/D k_0$ and $\mu_{\text{MF}} = \sqrt{(1-m_{\text{MF}}^2)/6}$. The corresponding energy is therefore given as 
$W^{(h)}_{0}
= 
- D k_0\left(m^2_{\text{MF}} +1 \right)/2$.

Finally, we calculate perturbatively the anisotropic contribution around the mean-field. For the anisotropy  of form \eqref{anisotropy1} (without spatial derivatives), the wave vectors of helices are not renormalized. Therefore, the contribution of anisotropy to the energy of the skyrmion lattice is calculated by using the same order parameter \eqref{skl braket}. Decoupling the ferromagnetic contribution $\vec m$ from Eq.\eqref{anisotropy2}, one obtains

\begin{equation}\begin{aligned}
\expect{{\cal H}_A} 
&= 
\frac{1}{2}A m^4+
A 
\expect{
\frac{1}{2}s_{x}^{4}+s_{y}^{4}+3s_{x}^{2}s_{z}^{2}+\frac{1}{2}s_{z}^{4}
}
\\
& +
A  m \expect{6 s_x^2 s_z+ 2 s_z^3}
+A m^2 \expect{3 s_x^2+3 s_z^2}.
\end{aligned} \end{equation}

\no
A direct calculation within the unperturbed state Eq.\eqref{skl braket} gives

\begin{equation}\begin{aligned}
\label{anistropy expect}
\expect{{\cal H}_A}
&= 
\frac{1}{2}A m^4+
\frac{27}{2}A m^2 \mu^2 
+ \frac{963}{32} A \mu^4
\\
&+
\frac{21}{\sqrt{2}}A \mu^3 m \cos(\varphi_1+\varphi_2+\varphi_3) .
\end{aligned} \end{equation}

\no
Note that the term containing $\mu^3$, is responsible for the minimization of the $3 Q$ structure, as was discussed in Section I. 
For $A>0$, $m>0$, $\mu>0$, expression 
\eqref{anistropy expect}
is minimized for $\varphi_1+\varphi_2+\varphi_3 = \pi$, which gives

\begin{align}
\label{anisotropy expect min}
\expect{{\cal H}_A}_0 = 
\frac{1}{2}A m^4+
\frac{27}{2}A m^2 \mu^2 
-\frac{21}{\sqrt{2}}  A \mu^3 m  
+ \frac{963}{32} A \mu^4 .
\end{align}

\no
The energy of the skyrmion lattice in the absence of electric field is thus given by \eqref{W0 in magnetic field} and \eqref{anisotropy expect min} in the mean-field approximation.

%
%
%
%

\section{Skyrmion Lattice in Electric Fields: Elastic distortion}

In this section we consider the shift in the SkL energy caused by  distortion of the skyrmion vectors. 
We start by re-writing the magneto-electric coupling in the rotated frame \eqref{EP coupling} in symmetrized form, which in units $D k_0$ is simply

\begin{equation}\begin{split}\begin{gathered}
\label{Hae}
\hat{\mathcal{H}}_{\ae} / D k_0= 2 \ae
\left( \begin{array}{ccc}
1 & \sqrt{2} & 0 \\
\sqrt{2} & 0 & 0 \\
0 & 0 & \text{-}1 \end{array} 
\right),
\end{gathered}\end{split}\end{equation}

\noindent
where we have introduced the {\textit{dimensionless electric field}} \ae, 

\begin{equation}\begin{split}\begin{gathered}
\ae \equiv \frac{\alpha E}{4 D k_0},
\end{gathered}\end{split}\end{equation}

\no
which plays the role of the small parameter of the theory.
To illustrate its smallness, we use typical electric fields $E=5 \times 10^{6} \, \text{V/m}$, and $\CuSe$ parameters as $J = 4.85 \times 10^{-23} \, \text{Jm} /\text{A}$, $k_0 = D/2 J = 10^8 \, \text{m}$, and ME coupling\cite{Mochizuki2015} is $\alpha \sim 10^{-14} \, \text{J} / \text{m}^{2} \text{V}$. This gives $\ae \sim 0.01$.
 Thus throughout this study, we build the perturbation theory in orders of $\ae^1$ and $\ae^2$, which is sufficient for describing both the symmetric and asymmetric responses in $E$-fields. 

First, we consider the helix vectors in external electric field.
Considering $E$-field as a small perturbation \eqref{Hae} on top of the $JD$-matrix \eqref{H0 2},  matrix perturbation theory gives\cite{Green1965}

%

\begin{equation}\begin{split}\begin{gathered}
\label{new eigenstate}
| {\vec S}^{(\ae)}_\ka   \rangle
=
| {\vec S}^{(0)}_\ka   \rangle
+\sum_{n \ne 0}
|{\vec S}^{(n)}_\ka   \rangle
\frac{
 \langle {\vec S}^{(n)}_\ka  |
\hat {\mathcal{H}}_{\ae}
| {\vec S}^{(0)}_\ka  \rangle
}
{\varepsilon^{(0)}_\ka-\varepsilon^{(n)}_{\ka}}
+ 
{\mathcal{O}} (\ae^2).
\end{gathered}\end{split}\end{equation}

\no
where $\varepsilon^{(n)}_{\ka}$ are given by Eq.\eqref{spectr} and  $| {\vec S}^{(n)}_\ka   \rangle$, $n=0,1,2$,  are eigenstates of $\mathcal H_0$ as given by Eqs.(14-16). 
The perturbed helix \eqref{new eigenstate} is by construction normalized on unity up to terms of order 
${\mathcal{O}} (\ae^2)$,

\begin{equation}
\label{constraint 1}
  \langle
 {\vec S}^{(\ae)}_\ka 
|
 {\vec S}^{(\ae)}_\ka 
 \rangle
=
1+ \mathcal{O}(\ae^2). 
\end{equation}

A direct calculation for the new helix, using formulas \eqref{new eigenstate}, \eqref{Hae}, \eqref{Sk0}-\eqref{Sk2},  gives

\begin{equation}\begin{split}\begin{gathered}
\label{elastic Sk}
| {\vec S}^{(\ae)}_\ka  \rangle 
=
| {\vec S}^{(0)}_\ka  \rangle
- 
\ae \, 
| {\vec F}_{\ka}  \rangle
+ 
{\mathcal{O}} (\ae^2),
\end{gathered}\end{split}\end{equation}

\no 
where ${\vec F}_{\ka}   = {\vec F} (\hat k_x, \hat k_y)\equiv {\vec F}_\ka(\phi)$ is the elastic distortion vector (here for each helix $\hat k_x = \cos \phi$, $\hat k_y = \sin \phi$, that is $\phi$ is angle between $\vec k$ and $\hat x$), so that

\begin{equation}\begin{split}\begin{gathered}
\label{Fk}
| {\vec F}_\ka \rangle
=
\left(
i F^x_{\ka} (\phi) ,
i F^y_{\ka} (\phi) ,
F^z_{\ka} (\phi) 
\right)^{\intercal} ,
\end{gathered}\end{split}\end{equation}

\no
with angle-dependent components

\begin{figure}[t]
	\includegraphics[width= 86.4581 mm ]{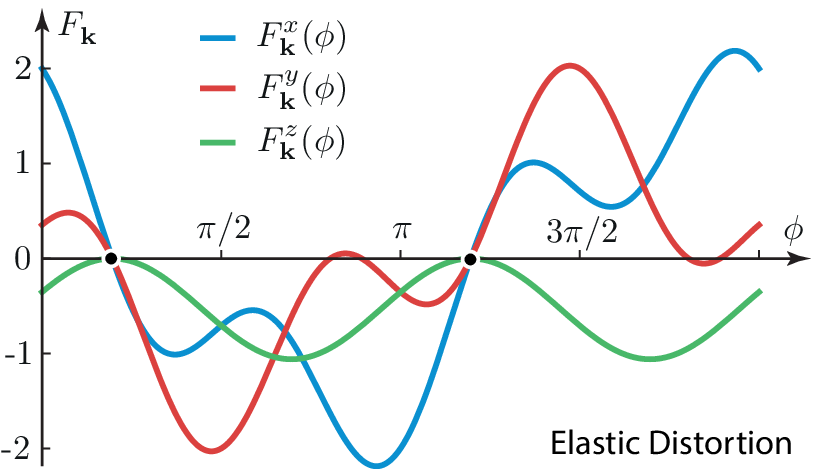}
	\caption{Components of elastic distortion vector 
	$ \vec F_{\ka}  $
	 as  functions of the helix angle $\phi$. $F_{\ka}^{x,y}$ are $2 \pi$-periodic while $F_{\ka}^{z}$ is $\pi$-periodic. The dots denote stationary points which are not effected by $E$-field.}
\end{figure}

\begin{align}
\label{Fkx}
F^x_{\ka} (\phi) 
&=
-\frac{\sin \phi }{\sqrt{2}}
+2 \cos ^3 \phi  
-\frac{3 \sin \phi  \cos ^2 \phi }{2 \sqrt{2}}
-\sin ^2 \phi  \,  \cos \phi ,
\\
\label{Fky}
 F^y_{\ka} (\phi)
 &=
\frac{\cos \phi }{2 \sqrt{2}}
-2 \sin ^3 \phi 
+\sin \phi  \cos ^2 \phi 
-\frac{3 \sin ^2 \phi \,  \cos \phi }{2 \sqrt{2}},
\\
\label{Fkz}
F^z_{\ka} (\phi)
& =
-\frac{1}{2 \sqrt{2}}
-\frac{\sin ^2 \phi  }{2 \sqrt{2}}
+\sin \phi  \cos \phi.
\end{align}

\no
The components $F^x_{\ka} $, $F^y_{\ka} $, $F^x_{\ka} $ of the elastic distortion vector are $\pi$ and $2\pi$ periodic functions, and are illustrated in Fig. 6. 
Note that the $x$ and $y$ components of both ${\vec S}^{(0)}_\ka$ and ${\vec S}^{(\ae)}_\ka$ are imaginary, and $z$ component is real  for both ${\vec S}^{(0)}_\ka$ and ${\vec S}^{(\ae)}_\ka$. 
Physically it means that the field-induced distortion of the skyrmion lattice does not change the orientation of the  skyrmion plane. 
One can verify both analytically and numerically that 
Eqs.\eqref{Fkx}-\eqref{Fky} together with \eqref{Fk} satisfy normalization \eqref{constraint 1}.








\section{Energy shift of the Skyrmion Lattice in Electric Fields}

In this section, we calculate the shift of the SkL energy in the first two orders in terms of dimensionless electric field $\ae$. 
This shift is contributed by the expectation value of the magnetoelectric coupling, and the anisotropic contributions due to the skyrmion lattice distortion. 

\subsection{Magneto-Electric Response}

The first contribution to the energy shift of the skyrmion lattice in electric field comes from taking the expectation value of \eqref{Hae} in the appropriate order in $\ae$. In this study, we consider $\ae$ to be small, thus the second order is sufficient for capturing the essential physics in both antisymmetric (field reversion $E \to -E$ gives the energy shifts of different signs) and symmetric (field reversion $E \to -E$ gives the energy shifts of same signs)   cases of interest. 
Using the explicit expression \eqref{elastic Sk}--\eqref{Fkz}, one obtains

\begin{equation}\begin{split}\begin{gathered}
\label{magn-el response}
\expect{ \hat{\mathcal{H}}_{\ae}}/Dk_0 = 
 -(2 m^2+3\mu^2) \, \ae - \frac{189}{4} \mu^2 \ae^2  + \mathcal{O} (\ae^3) .
\end{gathered}\end{split}\end{equation}

\no
The expression \eqref{magn-el response} is the leading contribution in case if the anisotropy of the system is small. Restoring the dimensional units, the shift of the SkL energy $W$ in electric field $E$ is therefore given by

\begin{equation}\begin{aligned}
\label{shift 1}
\Delta W_{1} (E)  = &  -  \frac{\alpha (2 m^2 + 3 \mu^2) M_s^2 a^3 }{4} E    
\\
& -   \frac{189\, \alpha^2 \mu^2 M_s^2 a^3  }{64 \, D k_0}  E^2 +  \mathcal{O} (E^3) .
\end{aligned}\end{equation}

\no
where $M_s$ is saturation magnetization (in $\text{A}/\text{m}$), $a$ is the lattice constant.
Note that the helimagnetic term $H_{JDh}$ is not giving shift contributions to this order.

\subsection{Anisotropy response}

Now, we calculate the direct contribution of anisotropy to the energy shift. The main physical mechanism here is the distortion of SkL in electric field, which results to perturbation in the anisotropic energy. To proceed, we explicitly use formulas  \eqref{elastic Sk}--\eqref{Fkz} to calculate the expectation value of the anisotropic term \label{anisotropy} in the new ground state. In the leading order one therefore obtains

\begin{equation}\begin{aligned}
\label{anis resp 1}
\expect{{\cal H}_A}_1 = 
\langle  \vec S^{(\ae)}      | \hat {\cal H}_A  |   \vec S^{(\ae)}   \rangle 
\simeq \frac{9}{8} A  m^2 \mu^2 \, \ae
 +\frac{99}{32} A \mu^4 \, \ae 
 \\
+\frac{81}{64} A  \mu ^4 \, \ae \, 
(
\cos 6 \phi_0
+
2 \sqrt{2} \sin 6 \phi_0
)
\\
+\frac{27  }{4 \sqrt{2}} A \, \ae \, m \mu ^3 \cos (\varphi_1+\varphi_2+\varphi_3)  .
\end{aligned}\end{equation}

\no
Here $\phi_0$ is the angle between the first skyrmion helix and $\hat x$ (note that due to $2 \pi/3$ symmetry in SkL rotation, and  $6 \phi$-arguments in Eq.\eqref{anis resp 1}, one can take  $\phi_0$ as the angle between any of the skyrmion helices and $\hat x$).
The same angular dependence  [third term in Eq.\eqref{anis resp 1}] were reported in study \cite{White2014}, where the minimization of SkL energy in electric fields leads to the SkL rotation in real space with respect to $\hat x$, if six-order anisotropy is considered, however all the angular-independent anisotropic contributions were there  neglected.\cite{White2014}
We notice that expression \eqref{anis resp 1} is dependent on the relative phases of the helices $\varphi_i$. The same situation occurs in  \eqref{anistropy expect}, when the mean-field energy is dependent on phase. We take again  $\cos(\varphi_1+\varphi_2+\varphi_3) = -1$, which minimizes the mean-field energy for fixed positive $m$, $\mu$, therefore

\begin{equation}\begin{aligned}
\label{anisotropic}
\expect{{\cal H}_A}_{1} & = 
\frac{9}{8} A \mu ^4  \left( 
\frac{11}{4}
- 3 \sqrt{2}  \frac{\mu}{m}
+
\frac{\mu^2}{m^2}
 \right)  \ae
 \\
& +\frac{81}{64} A  \mu ^4 
(
\cos 6 \phi_0
+
2 \sqrt{2} \sin 6 \phi_0
)
\, \ae  + \mathcal{O}(\ae^2).
\end{aligned}\end{equation}

\no
This contribution gives the linear anisotropic response to the external field. However, for the completeness of discussion, we need to take into account further terms $\mathcal{O}(\ae^2)$, which come both from the elastic distortion of the SkL and the inelastic (quadratic in $\ae$) distortion of the SkL.

\subsubsection{Mixed Inelastic response}

To consider the nonlinear anisotropic response, we re-define the perturbed helix eigenstates up to the second order, which are given by a perturbative expansion

\begin{equation} \begin{aligned}
\label{nonelastic distortion}
\ket{ {\vec S}^{(\ae)}_\ka  }
 =
\ket{ {\vec S}^{(0)}_\ka }
+\sum_{n \ne 0}
\ket{ {\vec S}^{(n)}_\ka  }
\frac{
\bra{ {\vec S}^{(n)}_\ka  }
\hat {\mathcal{H}}_{PE}
\ket{ {\vec S}^{(0)}_\ka }
}
{\varepsilon^{(0)}_\ka-\varepsilon^{(n)}_{\ka}}
\\
 +
\sum_{n,m \ne 0}
\ket{ {\vec S}^{(n)}_\ka  }
\frac{
 \bra{ {\vec S}^{(n)}_\ka  }
\hat {\mathcal{H}}_{\ae}
\ket{ {\vec S}^{(m)}_\ka }
 \bra{ {\vec S}^{(m)}_\ka  }
\hat {\mathcal{H}}_{\ae}
\ket{ {\vec S}^{(0)}_\ka }
}
{( \varepsilon^{(0)}_\ka-\varepsilon^{(n)}_{\ka}  )
\, ( \varepsilon^{(0)}_\ka-\varepsilon^{(m)}_{\ka} )
}
\\
 -
\sum_{n \ne 0}
\ket{ {\vec S}^{(n)}_\ka  }
\frac{
\bra{ {\vec S}^{(n)}_\ka  }
\hat {\mathcal{H}}_{\ae}
\ket{ {\vec S}^{(0)}_\ka }
\bra{ {\vec S}^{(0)}_\ka  }
\hat {\mathcal{H}}_{\ae}
\ket{ {\vec S}^{(0)}_\ka }
}
{
( \varepsilon^{(0)}_\ka-\varepsilon^{(n)}_{\ka}   )^2
}
\\
 -
\frac{1}{2}
\sum_{n \ne 0}
\ket{ {\vec S}^{(0)}_\ka  }
\frac{
\bra{ {\vec S}^{(0)}_\ka  }
\hat {\mathcal{H}}_{\ae}
\ket{ {\vec S}^{(n)}_\ka }
\bra{ {\vec S}^{(n)}_\ka  }
\hat {\mathcal{H}}_{\ae}
\ket{ {\vec S}^{(0)}_\ka }
}
{
( \varepsilon^{(0)}_\ka-\varepsilon^{(n)}_{\ka} )^2
}
\\
 + {\mathcal{O}} (\ae^3).
\end{aligned}
\end{equation}

\no
This expansion leads to re-definition of \eqref{elastic Sk} by adding a nonelastic distortion of the skyrmion lattice,

\begin{equation}\begin{split}\begin{gathered}
\label{2nd order basis}
| {\vec S}^{(\ae)}_\ka  \rangle
=
| {\vec S}^{(0)}_\ka  \rangle
- 
\ae \, 
| {\vec F}_{\ka}   \rangle
+ 
\ae^2 \, 
| {\vec G}_{\ka}  \rangle
+
{\mathcal{O}} (\ae^3),
\end{gathered}\end{split}\end{equation}

\no 
where ${\vec G}_{\ka}   = {\vec G} (\hat k_x, \hat k_y)\equiv {\vec G}_\ka(\phi)$ is the main inelastic distortion vector,

\begin{equation}\begin{split}\begin{gathered}
| {\vec G}_\ka \rangle
=
\left(
i G^x_{\ka} (\phi) ,
i G^y_{\ka} (\phi) ,
G^z_{\ka} (\phi) 
\right)^{T} . 
\end{gathered}\end{split}\end{equation}

\begin{figure}[b]
\label{inelastic}
	\includegraphics[width= 86.4581 mm ]{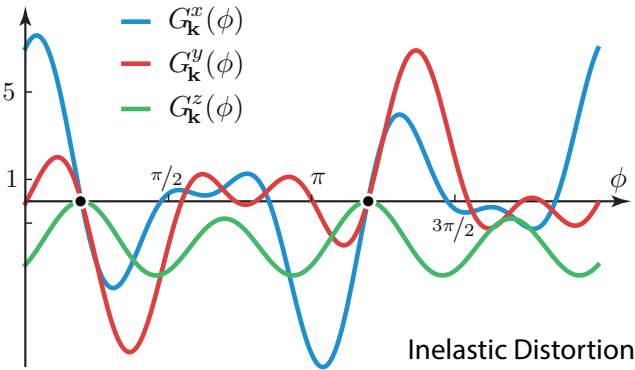}
	\caption{Components of the nonelastic distortion vector $\vec G_{\vec k}$ as a function of the helix direction angle $\phi$. $G_{\ka}^{x,y}$ are $2 \pi$-periodic while $G_{\ka}^{z}$ is $\pi$-periodic. The dots denote stationary points which are not effected by $E$-field.}
\end{figure}

\no
The direct calculation of the inelastic distortion vector, by using \eqref{nonelastic distortion}, \eqref{Hae}, \eqref{Sk0}-\eqref{Sk2}, gives

\begin{equation}\begin{aligned}
G^x_{\ka} (\phi)  & =
-\frac{39 }{64 \sqrt{2}} \sin \phi 
+\frac{131}{128 \sqrt{2}}  \sin 3 \phi 
+\frac{273 }{128 \sqrt{2}} \sin 5 \phi 
\\
&  +\frac{33 }{16} \cos \phi 
+\frac{119}{32} \cos 3 \phi
+\frac{39}{32} \cos 5 \phi  ,
\\
G^y_{\ka} (\phi) & =
-\frac{33 }{16} \sin \phi 
+\frac{37}{32} \sin 3 \phi 
+\frac{39}{32} \sin 5 \phi
\\
 & - \frac{93 }{64 \sqrt{2}} \cos \phi 
 +\frac{443 }{128 \sqrt{2}} \cos 3 \phi 
 -\frac{273 }{128 \sqrt{2}} \cos 5 \phi  ,
\\
G^z_{\ka} (\phi)
& =
\frac{3}{8} \sin 2 \phi 
+\frac{15}{16} \sin 4 \phi 
+\frac{3 }{16 \sqrt{2}} \cos 2 \phi 
\\
& -\frac{105 }{64 \sqrt{2}}  \cos 4 \phi 
-\frac{171}{64 \sqrt{2}}. 
\end{aligned}\end{equation}

\no
These dependencies are shown in Fig.6. The comparison between the elastic and inelastic distortion vectors is shown on Figure 7. Finally, for the numerical consistence of calculations, one can verify that the new ground state is normalized on unity,

\begin{equation}
 \langle
 {\vec S}^{(\ae)}_\ka 
|
 {\vec S}^{(\ae)}_\ka 
 \rangle
=
1+ \mathcal{O}(\ae^3). 
\end{equation}

\no
Therefore,  the mixed elastic-inelastic response $\mathcal{O}(\ae^2)$ is obtained by calculating the expectation value in the new basis \eqref{2nd order basis}.  The direct calculation gives

\begin{equation}
\label{mixed contribution}
\expect{{\cal H}_A}_{2} = 
\frac{27}{64} A \mu^4 
\left(
f_0+f_1 \cos 6 \phi_0 + f_2 \sin 6 \phi_0
\right)  
\ae^2 ,
\end{equation}

\no
where we have introduced the following dimensionless factors

\begin{align}
f_0 & = 
-\frac{191}{8}
+62 \sqrt{2} \, \frac{ m}{\mu }
-16  \frac{m^2}{\mu ^2} ,
\\
f_1 & = 
29
- 14 \sqrt{2} \, \frac{ m}{\mu }
+ 56 \frac{ m^2}{\mu ^2} ,
\\
f_2 & = 
-59 \sqrt{2}
+16 \frac{ m}{\mu }
-32 \sqrt{2} \, \frac{ m^2}{\mu ^2}.
\end{align}

\no
The mixed elastic-inelastic contribution \eqref{mixed contribution} gives the second-order correction in $\ae$ which may be important in some particular cases, for example, when the first order correction vanishes. Then, the electric field response does not depend on the field polarity ($\pm E$ is absorbed in $E^2$).

\begin{figure}
	\includegraphics[width= 86.4581 mm ]{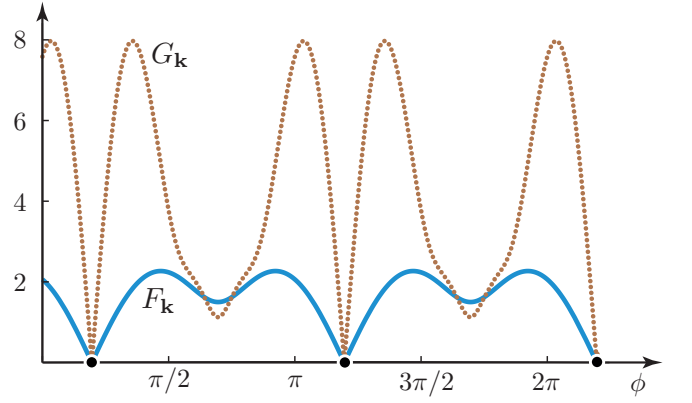}
	\caption{Comparison between magnitudes of elastic 
	$F_{\ka} =  \left| \vec F_{\ka}  \right|$
	 and nonelastic $ G_{\ka} = \left| \vec G_{\ka}  \right|$ distortion vectors as the function of the helix angle $\phi$. There $\pi$-periodic stationary points, which indicate the direction along which the helices are not disturbed, $F_k=G_k=0$.   }
\end{figure}

\

Therefore, we have three contributions to the shift in the SkL mean-field energy:

\begin{equation}\begin{aligned}
\label{dW1}
\Delta W_{1} (E)  =   -  \frac{\alpha (2 m^2 + 3 \mu^2) M_s^2 a^3 }{4} E    
 -   \frac{189\, \alpha^2 \mu^2 M_s^2 a^3  }{64 \, D k_0}  E^2 ,
\end{aligned}\end{equation}

\begin{equation}\begin{aligned}
\label{dW2}
\Delta W_2 (E)     = 
\frac{9 \alpha \, A \mu ^4 M_s^2 a^3 }{32 D k_0} E
\left[  
\frac{11}{4}
- 3 \sqrt{2}  \frac{\mu}{m} 
 +
\frac{\mu^2}{m^2}
\right.
\\ \left.
  +\frac{9}{8}
(
\cos 6 \phi_0
+
2 \sqrt{2} \sin 6 \phi_0
)
 \right] ,
\end{aligned}\end{equation}

\begin{equation}\begin{aligned}
\label{dW3}
\Delta W_3(E) & = 
\frac{27 \alpha^2 A \mu^4 M_s^2 a^3 }{1024 \,  D^2 k_0^2 }   
E^2
 \\
& \times \left(
f_0+f_1 \cos 6 \phi_0 + f_2 \sin 6 \phi_0
\right) .
\end{aligned} \end{equation}

\no
Together, formulas \eqref{dW1}-\eqref{dW3} give the contribution up to the second order in electric field $E$. In case of the weak anisotropy $A  \ll D^2/J$, the term \eqref{dW3} can be usually neglected. Note also that in the main-order approximation ($\ae^1$) the $E$-field induced shift in energy is also the shift in free energy of the skyrmion lattice for a fixed temperature near $T_C$.

It is therefore possible to stabilize (if $\Delta W_\text{total}<0$) or destabilize (if $\Delta W_\text{total}>0$) the skyrmion lattice by choosing the appropriate magnitudes and polarities of electric fields. This coincides with the previous idea of Mochizuki \cite{Mochizuki2016} of creating \textit{single} magnetic skyrmions with external $E$-fields in multiferroic  $\CuSe$.  The present study is therefore a bridge towards this idea in the bulk samples, where the skyrmions usually exist in the form of long-range-ordered or partially-disordered skyrmion arrays. The experimental research into writing and erasing skyrmions with electric fields in bulk $\CuSe$ is on the final stage in our lab and will be published elsewhere.

\section{Summary}

The present model describes the shift of the mean-field energy of the SkL, which is either positive or negative depending on the direction of the electric field. In a particular situation, when the first order terms  vanish, the energy shift is of the same sign for both field polarities.

An interesting output of the calculation is the existence of the stationary points if a helix is directed in a proper way (Fig. 7). This feature comes both in the elastic and nonelastic distortions of the skyrmion lattice. However, as the SkL is constructed on the three helices, the mean-field energy of the SkL is still shifted. 

Finally, we sketch the limitations of the calculation. First, this study describes the first two perturbative corrections to the mean-field energy of the skyrmion lattice in the multispiral approximation, without comparing the free energies of different possible phases (helical, conical) in the system. Second, the energy functional is taken in the quasiclassical continuous-field limit, which limits the use of the model only to the sufficiently low DMI parameter ($D/J \ll 1$) and excludes the quantum regime (low $T$). Third, the effect of electric field on critical fluctuations on top of the mean-field SkL solution is not considered as it comes as a higher-order contribution in the critical correlation length.

The main physical consequence of the phenomenon under study is the field-induced \textit{stabilization of the skyrmion phase in the bulk}, which has been so far indirectly observed in $\CuSe$. \cite{Okamura2016}
 This mechanism if further developed would allow one either to write or erase the skyrmion array over the full sample if it is properly placed in the $H$, $T$ phase diagram, and thus opens further routes for skyrmion-based racetrack logical elements and data storage devices.

\

\no
\textit{Acknowledgments}. - The work was supported by the Swiss National Science Foundation, its Sinergia network Mott Physics Beyond the Heisenberg Model (MPBH). The authors would like to thank  Achim Rosch and Jiadong Zang for useful discussion.

\end{document}